# From Street Form to Spatial Justice: Explaining Urban Exercise Inequality via a Triadic SHAP-Informed Framework

Minwei Zhao[a], Guosheng Yang[a], Zhuoni Zhang[a], Filip Biljecki[bc], Hanzhi Zu[a] and Cai Wu[a]*

**a** Thrust of Urban Governance and Design, The Hong Kong University of Science and Technology (Guangzhou), Guangzhou, China.

**b** Department of Architecture, National University of Singapore, Singapore

**c** Department of Real Estate, National University of Singapore, Singapore

## Abstract

Urban streets constitute essential everyday health infrastructure, yet their potential to support physical activity remains unevenly realized. Current assessments typically prioritize formal sports facilities or rely on coarse, area-level metrics, thereby failing to capture the micro-scale mechanisms of "exercise deprivation" within street networks. To address this gap, this study proposes a theory-informed, explainable analytical framework to diagnose street-level physical inactivity. We employ Henri Lefebvre's spatial triad as an interpretive lens, which conceptualizes urban space as a contradictory and co-productive unity composed of three interconnected dimensions: Conceived Space (representations of space), Perceived Space (spatial practice), and Lived Space (spaces of representation). Drawing on the research agenda of Critical Quantitative Geography, these interconnected dimensions are operationalized into measurable indicators derived from multi-source urban datasets. Subsequently, we employ SHapley Additive exPlanations (SHAP) to quantify the non-linear contributions of these dimensions to exercise intensity.

Using Shenzhen as a case study, our analysis reveals that while Conceived attributes explain the largest share of variance globally, local deprivation mechanisms exhibit significant heterogeneity across development contexts. By decomposing these driving forces, we identify a seven-mode typology of deprivation, which classifies street segments based on specific spatial deficits (e.g., "Perceived-only deprivation" in structurally sound but visually hostile environments). Furthermore, a bivariate spatial association analysis identifies clusters of "high-demand/low-support" mismatch, pinpointing priority areas for intervention. Methodologically, this study reconciles the predictive power of machine learning with the interpretive depth of spatial theory; practically, it transforms abstract spatial concepts into a segment-scale, diagnostic tool for advancing spatial justice in everyday physical activity.

**Keywords:** Street-based physical activity; Exercise deprivation; Explainable machine learning; Street-view imagery; Social sensing; Spatial mismatch; Spatial justice.

## 1 Introduction

Physical inactivity is globally recognized as a major risk factor for noncommunicable diseases, obesity, mental health issues, and premature mortality (Sallis et al., 2016). As cities become denser and more complex, ensuring equitable access to opportunities for physical activity has become a pressing concern for urban planning and public health (Kohl et al., 2012; World Health Organization, 2019). Traditionally, research on active living has prioritized public parks, greenways, and designated sports facilities as primary sites for promoting physical activity (Kaczynski & Henderson, 2008). Such studies often rely on coarse measures of accessibility, typically treating opportunities as discrete, point-based destinations (Schipperijn et al., 2017). However, recent scholarship increasingly identifies urban streets as crucial yet underutilized spaces for everyday activity (Thompson, 2013; Kostrzewska, 2017). Unlike formal facilities, streets are ubiquitous and low-barrier, offering continuous environments for walking, jogging, or cycling, particularly in dense and underserved areas with limited designated infrastructure (You, 2016).

However, current approaches to assessing street-level supportiveness remain methodologically limited. Conventional indicators often overlook the continuity of streets and their capacity to support diverse and spontaneous forms of physical activity. Moreover, mainstream assessments typically rely on aggregated "walkability indices," which operate as a form of analytical "black box." By conflating structural connectivity with environmental quality, such indices obscure the specific drivers of street-level activity. These coarse diagnostics fail to capture the micro-mechanisms underlying "exercise deprivation": they may identify where streets fail to support activity, but not why. For instance, a street may be structurally accessible yet functionally "deprived" due to poor sensory environments or a lack of perceived social safety. Understanding streets not merely as conduits for movement but as spaces that shape everyday behavior therefore requires attention to their inherent complexity.

Critical geographer Henri Lefebvre confronted precisely this elusive yet powerful nature of space (Aristotle, 1982). The epistemological foundation of his theory lies in the spatial triad, which articulates three interrelated dimensions of space in order to illuminate the complexities of everyday life (Watkins, 2005). Lefebvre argues that space is fundamental to lived experience and that each experience encompasses three mutually constitutive aspects: representations of space (conceived space), spatial practices (perceived space), and spaces of representation (lived space) (Lefebvre, 1991).

Although this triadic framework has been widely employed in studies of spatial justice and inequality and has become a cornerstone of critical geography, its application in empirical research has long been characterized by an ontological–epistemological–methodological tension. Lefebvre's spatial ontology resists fragmenting space into discrete and measurable variables, whereas quantitative urban analysis relies on operationalized indicators to decompose spatial mechanisms. Developments in Critical Quantitative Geography have sought to move beyond this binary opposition between dialectics and quantification (Kwan, M. P., & Schwanen, T., 2009a) by advocating a post-cultural turn perspective on quantitative approaches (Kwan, M. P., & Schwanen, T., 2009b). At the same time, advances in quantitative methodologies have provided a practical methodological "toolbox" for fostering interaction between these traditions.

To address these gaps, this study proposes a theoretically grounded and interpretable analytical framework for diagnosing insufficient physical activity at the street level. We employ Lefebvre's spatial triad—conceived space (C), perceived space (P), and lived space (L)—as an interpretive lens to examine the complex interplay among institutional design, sensory environments, and patterns of use. Within this framework, conceived space captures the rationalities embedded in street hierarchies, connectivity structures, and land-use planning; perceived space encompasses the visual and tactile cues that shape how space is interpreted and navigated; and lived space reflects the situated, symbolic, and affective engagements manifested through presence and activity. By integrating this theoretically informed perspective with explainable machine learning—specifically SHapley Additive exPlanations (SHAP)—we aim to reveal association patterns that are both analytically legible for urban research and meaningful for policy reflection, while avoiding the reduction of urban spatial complexity to overly simplistic indicators.

Consequently, this study makes four primary contributions to the literature on spatial justice and active living:

(1) We introduce a triad-informed framework for indicator construction, employing Lefebvre's dimensions as an organizing lens to structure diverse urban data. This approach moves beyond ad hoc feature stacking, offering a theoretically grounded method for categorizing the built environment using indicators from road networks, street-view imagery, and social sensing.

(2) We link these spatial factors to actual street-level exercise behaviors using explainable machine learning. This reconciles the predictive power of machine learning with the interpretive depth of spatial theory, enabling us to not only identify which spatial attributes are most strongly associated with physical activity but also to interpret these relationships in a transparent manner.

(3) We develop a novel seven-mode typology of deprivation (e.g., Perceived-only deprivation vs. Triple deprivation). Unlike traditional aggregate metrics that treat inequality as a simple deficit of resources, this typology conceptually reframes deprivation: it demonstrates that spatial failure arises not merely from the absolute absence of infrastructure, but from the relational misalignment between physical form, sensory quality, and social practice. This offers a granular diagnostic lens, revealing how even structurally well-connected streets may be functionally deprived due to specific sensory or social imbalances.

(4) Finally, our framework enables the identification and contextualization of spatial mismatches through the lens of our triad-informed deprivation typology. By cross-referencing mismatch clusters with dominant deprivation profiles, we identify where targeted design or policy responses can be most effectively directed toward context-sensitive improvements in everyday movement support.

Together, these contributions help reconcile data-driven urban analytics with spatial theory by drawing on its structural insights to inform how we organize and interpret multi-source indicators. While our approach remains empirically grounded, the triadic orientation offers a flexible structure for reasoning about spatial inequality across conceived, perceived, and lived dimensions. This synthesis enables a more legible and context-aware diagnosis of exercise deprivation, with potential implications for targeted urban interventions across diverse settings.

# 2 Literature review

## 2.1 Urban Physical Activity and Exercise Deprivation

A substantial body of research on urban physical activity has focused on identifying how formal designated sports amenities, such as parks, greenways, and purpose-built sports facilities, provide focal nodes for exercise (Kaczynski & Henderson, 2008; Duan et al., 2018; Zhang et al., 2024). Within this paradigm, physical activity opportunities are often operationalized through accessibility measures and proximity-based indices that treat destinations as discrete point resources (Schipperijn et al., 2017), while concerns about physical inactivity have motivated public health scholarship emphasizing the uneven distribution of these facilities across populations (Sallis et al., 2016; Kohl et al., 2012; World Health Organization, 2019). Frameworks such as “activity deserts” (Coombes et al., 2010) and related accessibility indices (Pate et al., 2021) thus primarily conceptualize inequality through the availability and spatial allocation of designated infrastructures. Promoting equity in physical activity is therefore examined largely through the lens of facility provision rather than through the more granular mechanics of everyday urban environments. Yet, emerging scholarship increasingly identifies urban streets as crucial yet underutilized spaces for everyday activity (Thompson, 2013; Kostrzewska, 2017). Recent studies emphasize that streets function as everyday, low-threshold public environments where diverse forms of movement unfold (Chen et al., 2024; Mehta, 2008; You, 2016). Their importance becomes especially evident in dense or underserved urban neighborhoods where formal recreational infrastructures are scarce. In such contexts, streets often become the default realm for everyday exercise among populations with limited access to designated amenities, a pattern that is particularly pronounced in high-density cities such as Shenzhen (Sallis et al., 2020; Chen et al., 2016).

This shift in attention from formal spaces to street environments calls for a parallel rethinking of how deprivation is conceptualized. The notion of "exercise deprivation" moves beyond infrastructure absence to encompass relational, multi-dimensional barriers embedded in street-level conditions: ranging from missing sidewalks to visual discomfort, safety concerns, or limited patterns of use and engagement (Ewing & Handy, 2009; Gehl, 2013; Alfonzo, 2005). However, prevailing approaches often rely on grid-based metrics or node-based proximity, which overlook the continuous character, perceptual attributes, and localised usage patterns of streets (Gehl, 2013). From a micro-scale perspective, a street may be "deprived" not only because of material deficits but also due to perceived barriers and limited social engagement or utilization (Alfonzo, 2005; Handy et al., 2002). These barriers often manifest unevenly, varying significantly from one street segment to the next. Understanding deprivation through this lens foregrounds the interplay of physical form, perception, and lived experience in shaping spatial inequality.

In sum, existing conceptual and analytical frameworks are not yet sufficient to characterize the complexity of street-based exercise deprivation. While shifting the focus to streets is crucial, current metrics often lack the structural depth to capture the "relational barriers" described above. There remains a critical need for a comprehensive framework that moves beyond ad hoc feature selection to structure multi-source data into coherent dimensions—planning, perception, and practice. This gap necessitates the triad-informed indicator construction developed in this study, which seeks to capture the layered nature of street space.

### 2.2 Spatial Data-Driven Approaches for physical activity equity analytics

Recent research has seen rapid advancements in data-driven analytical methods. Geographic information science, computer vision, mobile sensing, and big urban data have enabled unprecedented granularity in modeling urban representations (Wu et al., 2023; Wu et al., 2024; Zhao et al., 2025). Researchers routinely integrate diverse spatial data to construct comprehensive representations of urban environments, an approach that has become central to contemporary urban health and human-environment interaction studies (Goodchild, 2007; Batty, 2013).

However, the surge in technical feasibility and predictive accuracy has often come at the expense of meaningful theoretical frameworks that link spatial form explicitly to lived urban experience (Kim, 1999; Connaway et al., 2011). The abundance of heterogeneous data from greenery indices to pedestrian reviews, frequently introduces challenges such as multicollinearity and inconsistent scaling (Zhang et al., 2019; Zhong et al., 2020). More critically, composite metrics like "walkability indices" often aggregate numerous variables into single scores without elucidating how individual indicators interact (Frank et al., 2010; Maghelal & Capp, 2011). Similarly, while POI density serves as a proxy for functional richness, it is seldom interpreted within a broader socio-behavioral context, limiting its diagnostic utility (Yue et al., 2017; Talen, 2003). Consequently, many urban analysis models function as "black-box" systems: proficient at forecasting phenomena yet inadequate in delivering actionable explanations or policy implications (Argyris, 1996). These predominantly top-down perspectives rarely explicitly incorporate the reciprocal interactions between spatial form, human perception, and lived social practices (Tang et al., 2020), limiting the understanding of exercise environments as dynamically experienced spaces.

To mitigate this interpretability challenge, explainable machine learning (XAI) methods, notably SHAP (SHapley Additive exPlanations), have emerged to quantify feature contributions and capture localized spatial variations (Lundberg et al., 2018; Molnar et al., 2020; Guidotti et al., 2019). However, a critical gap remains: existing applications of SHAP in urban research are largely descriptive, serving as post-hoc tools to rank feature importance without integration into broader theoretical frameworks (Samek et al., 2021). Without a guiding theory, SHAP values explain the mathematical weight of a feature, but not necessarily its spatial or social significance.

This theoretical deficit also compromises the spatial resolution of analysis. Lacking a framework grounded in experiential space, researchers often rely on coarse, grid-based units, which obscure micro-scale inequalities and local discontinuities, particularly in fragmented urban typologies (Gavrilidis et al., 2022). Although recent efforts attempt to employ finer units such as streetscape segments (Fang et al., 2025), these often remain statistical constructs rather than theoretically grounded units of lived experience (Gehl, 2013; Lynch, 1960). Consequently, there is a pressing need for a framework that not only utilizes granular units and explainable algorithms but anchors them within a robust spatial theory.

In sum, despite the granular capabilities of contemporary urban analytics, the field lacks a unifying framework that bridges algorithmic prediction with the lived reality of street space. Advancing equity analytics requires moving beyond "black-box" modeling toward conceptually anchored diagnostics. To bridge this divide, there is a pressing need for methods that reconcile the high predictive power of machine learning with the interpretive depth of critical-geographical theory. This motivates our use of explainable AI (SHAP) to decode the specific spatial drivers of inactivity, transforming opaque algorithms into transparent policy evidence.

### 2.3 Triadic Readings of Urban Space

Henri Lefebvre's spatial triad remains one of the most influential frameworks for interpreting the layered production of urban space (Simonsen, 2005). In The Production of Space (1974), Lefebvre conceptualized space through three interrelated dimensions: conceived space (formal representations and planning rationalities), perceived space (sensory cues and everyday practices), and lived space (symbolic meanings and embodied experiences). This perspective has been widely mobilized across geography, sociology, planning, and architecture to examine how inequalities are produced not only through infrastructure but also through perception and practice (Soja, 2010; Harvey, 2006). In urban health research, it is particularly valuable because it shows how exclusion may stem as much from perceptual and experiential conditions as from material deficits (McLees, 2013; Pierce & Lawhon, 2015).

Applied to street-based exercise, the triad provides an interpretive orientation for understanding why environments that appear similar in form may diverge in their capacity to support movement. Two streets with comparable width or connectivity, for instance, may yield quite different levels of activity depending on how safe they feel, how socially engaging they are, or how comfortable they appear (Handy et al., 2005). Such divergence reflects not only material affordances but also relations of visibility, power, and normativity that shape those who feel entitled to linger, pass, or exercise (Wacquant, 2022). In this sense, exercise deprivation is relational and multidimensional: it arises from the alignment, or misalignment, of planning logics, perceptual cues, and lived practices.

From a geographical perspective, the triad also sharpens long-standing debates about how planned orders meet everyday tactics and negotiations where conceived intentions are appropriated, subverted, or re-made in practice (de Certeau, 1984). Perceived space involves more than visual legibility; it comprises atmospheres, affordances, and embodied judgements that mediate movement and safety (Anderson, 2009). Lived space is relational and plural, constituted by intersecting trajectories rather than a single, coherent narrative (Massey, 2005). These insights matter for street-based exercise: the same formal configuration can host distinct rhythms of use, confidence, and co-presence as conditions shift across groups and times of day (Edensor, 2012).

In bridging these theoretical insights with empirical analysis, we adopt the triad not as a rigid taxonomy, but as a flexible reading. In this framework, Conceived space foregrounds institutional form and hierarchy; Perceived space attends to sensory and visual conditions; and Lived space traces rhythms of presence and co-activity. Crucially, we recognize that boundaries between these dimensions remain porous. A feature like street lighting, for instance, operates simultaneously as a perceptual signifier (atmosphere) and an organizational cue embedded in everyday routines (structure). Rather than viewing these overlaps as methodological flaws, we treat them as analytically generative, acknowledging the hybrid nature of urban settings.

Taken together, these dimensions offer a structured yet adaptable vocabulary for reviewing how movement takes shape along streets. The intersection of these literatures highlights a clear analytical imperative: the need for a street-scale diagnosis that is empirically tractable yet conceptually anchored to this "adaptable vocabulary." While emerging data-driven approaches offer high-resolution evidence, they often lack the theoretical scaffolding to explain why supply fails to meet demand in specific contexts. By integrating these dimensions with XAI, we pursue a segment-scale, relational account that crystallizes modeled outcomes into a diagnostic typology. Crucially, by contextualizing these spatial profiles against population demand, we aim to demonstrate that the "mismatch" between infrastructure and usage is rarely a simple matter of physical absence or low accessibility, but rather a symptom of incoherence among the conceived, perceived, and lived dimensions. Ultimately, this approach shifts the analytical focus from simple prediction to structural diagnosis, providing a robust vocabulary for interpreting the spatial justice of everyday movement.

## 3 Methodology

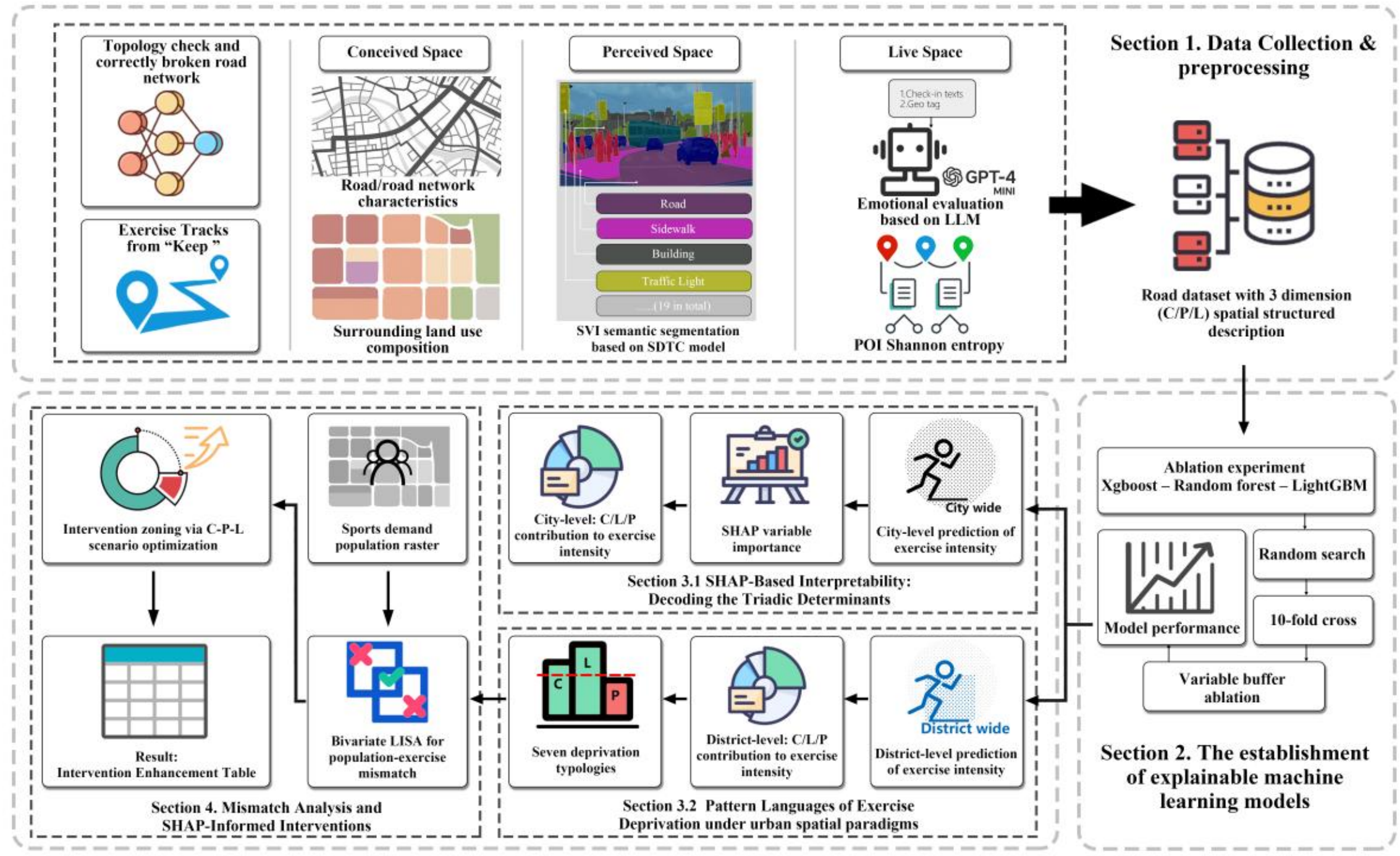

**Figure 1 Analytical Workflow: A Theory-Informed, Explainable Pipeline. (1) Data Integration: Aligning multi-source data to street segments to construct C/P/L descriptors. (2) Predictive Modelling: Training gradient-boosted trees (XGBoost) with cross-validation to capture non-linear spatial associations. (3) SHAP-Based Interpretation: Decomposing predictions into triadic determinants at citywide and district scales (4.2), and synthesizing these contributions into a seven-mode diagnostic typology of deprivation (4.3). (4) Mismatch & Intervention: Identifying spatial mismatches between population demand and street supportiveness to guide targeted urban renewal.**

To diagnose street-level exercise deprivation in a spatially structured and interpretable manner, we developed a comprehensive analytical framework (Figure 1). Addressing the gap between "black-box" algorithms and spatial theory identified in the literature, this workflow bridges abstract spatial concepts and actionable planning evidence through a four-stage logical progression. First, we organize multi-source urban big data into a triad-informed indicator system, employing Lefebvre’s dimensions as an interpretive lens to structure measurable street-level attributes. Second, we employ explainable machine learning to quantify the non-linear contributions of these dimensions to exercise intensity. Third, based on these attributions, we synthesize a diagnostic typology to characterize the specific nature of deprivation. Finally, we integrate this typology with a grid-based mismatch analysis to target interventions where demographic demand exceeds spatial support.

**Figure 2. Spatial distribution of street-segment exercise intensity in Shenzhen. The inset map shows the location of Shenzhen within the Greater Bay Area. The main panel presents the log-transformed KEEP intensity at the street-segment level across Shenzhen. Panels (a – d) provide enlarged views of four representative urban areas highlighted in the main map, illustrating local heterogeneity in exercise intensity across different street networks and built-up contexts.**

We situate our analysis in Shenzhen, a rapidly growing megacity in southern China (Figure 2). As a frontier of varying urban experiments, Shenzhen provides a compelling laboratory to examine the spatial production of exercise opportunities. Its juxtaposition of rapid, top-down master planning (Conceived space) with vibrant, often informal bottom-up activity (Lived space) creates a complex landscape ideal for diagnosing spatial deprivation. Uniquely, the city possesses high-resolution, multi-source urban data, enabling a fine-grained empirical translation of the conceived, perceived, and lived representations central to our framework. To ensure analytical consistency, all datasets were aligned to the road segment level and underwent rigorous topological preprocessing (Table 1).

**Table 1 Data sources and description table**

| | **Data Source** | **Description** | **Variable** |
|---|---|---|---|
| **Exercise Trajectory** | KEEP 2019–2021 exercise trajectory data (KEEP Inc., 2025) | Individual-level GPS trajectories represent jogging, walking and cycling behavior. | 'log_d10_norm', 'log_d20_norm', 'log_d30_norm' |
| **Conceived Space** | 2024 Amap detailed road network (Amap, 2025) | High-resolution road network dataset containing road type, hierarchy, connectivity, and functional attributes. | C_free_speed； C_capacity … C_ deg 800m; C_betw_800m C_clo_800m; C_depth_800m |
| | 1m resolution land use remote sensing (Li et al., 2023) | Landuse classification raster providing fine-scale spatial composition around roads, supporting objective environmental context analysis. | C_D_transport; C_D_tree … C_D_water; C_D_building |
| **Perceived Space** | Baidu Street View 2024 segmentation results based on STDC model (Fan et al., 2021) | Street-level semantic features extracted from 2024 Baidu Street View using the state-of-the-art STDC segmentation model to represent visual perception. | P_road; P_sidewalk … P_vegetation; P_sky |
| **Lived** | Weibo check-in | Spatiotemporal social media | L_total_weibo_count |

| **Space** | data 2021-2025 (Sina Weibo, 2025) | data capturing public expression and spatial behavior. | L_sport_mean; L_urban_mean L_positive_mean L_context_mean |
| --- | --- | --- | --- |
| | Amap 2021 POI (Amap, 2025) | Point-of-interest data offering categorical diversity and entropy, representing functional heterogeneity around streets. | L_poi_entropy300 |
| **Mismatch analysis** | 2022 90m resolution population grid data (Lebakula et al., 2025) | Gridded population data with 90m resolution reflecting spatial distribution of urban residents, used for mismatch diagnosis with exercise intensity. | Grid population |

**Figure 3 Representative spatial distributions of street-level indicators across the conceived, perceived, and lived dimensions.**

The analysis centers on street-level physical activity intensity as the dependent variable, representing the observed outcome of spatial interaction. This metric was derived from anonymized GPS trajectories recorded by the Keep fitness platform (Keep Inc., 2025), capturing jogging, walking, and cycling behaviors. For each road segment, intersecting trajectory points were counted and normalized by segment length, then log-transformed and z-score standardized to produce robust exercise density metrics under three distance bands (10 m, 20 m, and 30 m). The resulting spatial pattern of exercise intensity across Shenzhen is shown in Figure 2.

To operationalize the explanatory dimensions of conceived, perceived, and lived space, we constructed a multi-source indicator system at the street-segment level. Figure 3 presents six representative indicators selected from these three dimensions, highlighting both their citywide spatial distributions and their neighborhood-scale heterogeneity through enlarged local views.

Conceived space, reflecting the "representations of space" and planning rationalities, was characterized through the structural skeleton of the street network and its functional context (Figure 3a). We constructed this description using two primary sources. First, a 2024 high-resolution road network dataset from Amap (Amap, 2025) provided intrinsic metadata including free-flow speed, design capacity, and functional classification. To capture topological hierarchy, network geometries were corrected to ensure graph validity, allowing the calculation of global and local centrality measures (degree, closeness, betweenness) within 800-meter ego-graphs, a scale consistent with the widely recognized 10-minute neighborhood framework (Staricco, 2022). Complementing this network logic, we incorporated 1-meter resolution land use classification data (Li et al., 2023) to compute the surrounding functional composition within 100-meter buffers, delineating the planned context through proportions of residential, commercial, transportation, vegetation, water, and building classes.

Perceived space corresponds to the sensory deciphering of the environment, representing the visual interface encountered by pedestrians. We characterized this dimension by applying semantic segmentation to 2024 Baidu Street View imagery using the state-of-the-art STDC model (Fan et al., 2019). Frontal and lateral views were sampled along street networks and parsed into visual categories such as sky, vegetation, sidewalk, façade, and vehicle (Figure 3b). These aggregated visual proportions were normalized by road length to form a perceptual profile for each segment, reflecting key experiential qualities such as visual enclosure, greenery, and the legibility of urban form.

Lived space embodies the "representational spaces"—the complex, symbolic, and affective reality overlaid on the physical form. Since this dimension reflects dynamic rhythms of practice rather than static objects, we approximated it through two digital proxies. First, over 1.5 million Weibo check-in posts (2021–2025) were analyzed using a ChatGPT-based semantic scoring approach with a structured prompt (see *Appendix A*). This approach scored each post across nine experiential dimensions, including sociability, sport-relatedness, and real-time context. Second, we calculated POI Shannon entropy from 2021 Amap POI data (Amap, 2025) to represent the diversity of surrounding amenities. Together, these indicators serve as a proxy for informal engagement, spatial vibrancy, and the potential for spontaneous activity that unfolds beyond formal planning (Figure 3c).

While our primary predictive modeling operates at the street segment level, assessing supply-demand mismatches requires integrating demographic data available only in raster format. We therefore rasterized the study area into a uniform 200 × 200m grid to bridge these spatial units. Population estimates from 2022 90-meter raster data (Lebakula et al., 2025) were aggregated into this new grid using area-weighted interpolation. Simultaneously, street-level C/P/L features and exercise variables were aggregated using length-weighted means within each grid cell. Finally, all continuous variables across the workflow (except categorical descriptors) were log-transformed and z-score normalized to ensure statistical comparability and suitability for gradient-based machine learning.

### 3.1 Quantifying Spatial Drivers: Predictive Modeling of Non-linear Associations

With the triad-informed dataset established, we proceeded to the computational phase to quantify how these spatial dimensions drive street-level activity. Four regression models are deployed to estimate the non-linear relationships between spatial features and exercise intensity, namely Ordinary Least Squares, Random Forest, LightGBM, and XGBoost, on the preprocessed dataset. Hyperparameter tuning was performed via randomized search combined with 10-fold cross-validation to ensure out-of-sample generalizability. The parameter space and final selected hyperparameters for each model are detailed in *Appendix C*.

### 3.2 From Attribution to Diagnosis: SHAP-Based Interpretation and Deprivation Typology

Building on the best-performing model, we applied SHapley Additive exPlanations (SHAP) to decompose predictions into additive contributions of input features. This interpretation was conducted at two scales: a citywide reading that summarizes contributions into the three canonical dimensions (Conceived, Perceived, Lived), and a district-level refinement where the conceived dimension was further disaggregated into structural ( $C_{\text{structure}}$ ) and functional ( $C_{landuse}$ ) components to capture local heterogeneity in planning legacies. This summary offers a compact, theory-attuned reading of the model output, relating to structural form (Conceived), streetscape conditions (Perceived), and patterns of presence (Lived) to the modelled outcome, while acknowledging overlaps across dimensions. To aid interpretation and pattern recognition, we visualized SHAP results using ranked importances and SHAP decision trajectories, stratifying segments by predicted deprivation severity and tracing principal contributions across the C/P/L readings.

Subsequently, we translated these SHAP attributions into a diagnostic typology of deprivation. Moving beyond unidimensional rankings, we implemented a rule-based classification to identify the dominant "deficit driver" for underperforming segments. By applying a relative threshold (e.g., the top 20% of negative SHAP impacts), segments were classified into seven mutually exclusive types: single-mode deprivation (C-only, P-only, L-only), dual-mode deprivation, or triple deprivation (CPL). This typology transforms abstract coefficients into a structural diagnosis that characterizes the specific configuration of constraints inhibiting physical activity, identifying whether the barrier manifests as an isolated deficit or a systemic mismatch among planning, perception, and practice. Formally, a segment s with normalized SHAP scores $\varphi_C$, $\varphi_P$, $\varphi_L$ is classified as:

$$Type(s) = \begin{cases} \boldsymbol{C-only} & \varphi_C > q_{0.8}, \varphi_P, \varphi_C \leq q_{0.8} \\ \boldsymbol{CP} & \varphi_C, \varphi_P > q_{0.8}, \varphi_C \leq q_{0.8} \\ & \cdots \\ \boldsymbol{CPL} & \varphi_C, \varphi_P, \varphi_L > q_{0.8} \end{cases}$$

### 3.3 From Diagnosis to Intervention: Spatial Mismatch Analysis

Finally, to address the scalar mismatch between linear street networks (supply) and areal population distribution (demand), we conducted a Bivariate Mismatch Analysis. Using the 200-meter grid integrated in the data construction phase, we applied Local Indicators of Spatial Association (LISA) to detect localized clusters where high population density coincides with low predicted exercise supportiveness. These "High-Demand/Low-Support" hotspots were then cross-referenced with the segment-level diagnostic typology. This synthesis bridges the gap between data analytics and intervention: the grid analysis identifies where the intervention is most urgently needed based on demographic pressure, while the typology informs what specific spatial dimension requires remediation.

This integrative framework thus combines high-performing prediction, theoretical attribution, spatial typology, and design-guided simulation to bridge data and intervention; providing actionable insights into how, where, and why urban streets fall short in supporting equitable physical activity.

## 4 Analyses and Discussion

### 4.1 Model Performance and the Necessity of Non-linear Approaches

To establish a robust analytical baseline for the triadic diagnosis, we evaluated the predictive capacity of four distinct regression algorithms, ranging from traditional Ordinary Least Squares (OLS) to advanced ensemble methods including Random Forest, LightGBM, and XGBoost. We employed a rigorous 10-fold cross-validation protocol to assess out-of-sample performance under randomized hyperparameter configurations, using the coefficient of determination ($R^2$) and root mean squared error (RMSE) as primary benchmarks.

The comparative results detailed in Table 2 reveal a significant divergence in model efficacy that validates our theoretical premise regarding urban complexity. The linear regression model accounted for only 28% of the variance ($R^2 = 0.2800$), a limited performance that underscores the inherent non-linearity of street-level exercise determinants which traditional linear assumptions fail to capture. In contrast, the tree-based ensemble methods demonstrated substantially higher predictive power. Notably, the XGBoost model achieved the best overall performance, attaining a mean cross-validated $R^2$ of 0.7023 (±0.0043) and the lowest RMSE of 0.0449 (±0.0003). This demonstrated superiority in both accuracy and stability confirms that capturing complex, non-linear spatial interactions is critical for adequately modeling urban activity. Consequently, XGBoost was selected as the reliable foundation for the subsequent SHAP-based attribution and the construction of the spatial typology.

**Table 2 Cross-validated model performance comparison across regression algorithms (mean ± standard deviation over 10 folds).**

| Model | CV R² (Mean ± Std) | CV RMSE (Mean ± Std) |
| --- | --- | --- |
| XGBoost | 0.7023 ± 0.0043 | 0.0449 ± 0.0003 |
| RandomForest | 0.3995 ± 0.0043 | 0.0638 ± 0.0005 |
| LightGBM | 0.5661 ± 0.0047 | 0.0542 ± 0.0004 |
| LinearRegression | 0.2800 ± 0.0137 | 0.0699 ± 0.0010 |

## 4.2 Decoding the Triadic Determinants: Multi-Scalar SHAP Interpretation

### 4.2.1 Global Hierarchy and Non-linear Mechanisms

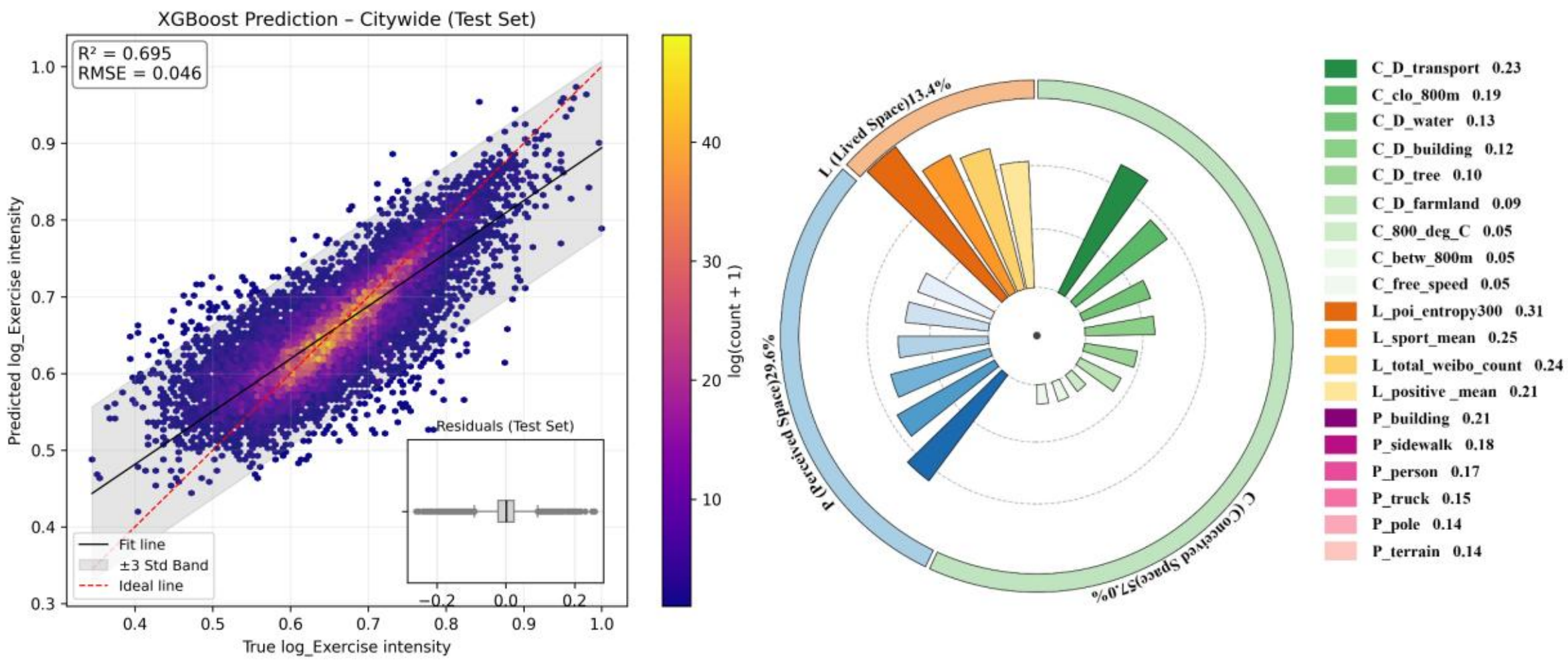

**Figure 4 Citywide model fit and triadic SHAP contribution overview.**

To unpack the spatial drivers of exercise intensity, we first examined the model's performance and feature attribution at the citywide scale. Figure 4 presents the global prediction performance alongside a triadic attribution analysis. The scatter plot demonstrates a strong agreement between predicted and observed exercise intensity ($R^2$ = 0.695), with residuals largely contained within the $3\sigma$ band, indicating a robust fit across the city's diverse urban fabrics. Decomposing these predictions via SHAP grouping reveals a distinct hierarchy of influence where the Conceived perspective accounts for the majority of explained variance (57.0%), followed by the Perceived perspective (29.6%) and the Lived perspective (13.4%). This hierarchy underscores that macro-level planning variables, such as transport infrastructure and land use, remain the dominant drivers of exercise potential. These findings echo urban design literature emphasizing that movement infrastructure constrains and conditions the fundamental opportunity spaces for physical activity (Handy et al., 2002; Cervero & Kockelman, 1997). While sensory and social qualities are critical, they operate within the structural envelope defined by the conceived skeleton of the city.

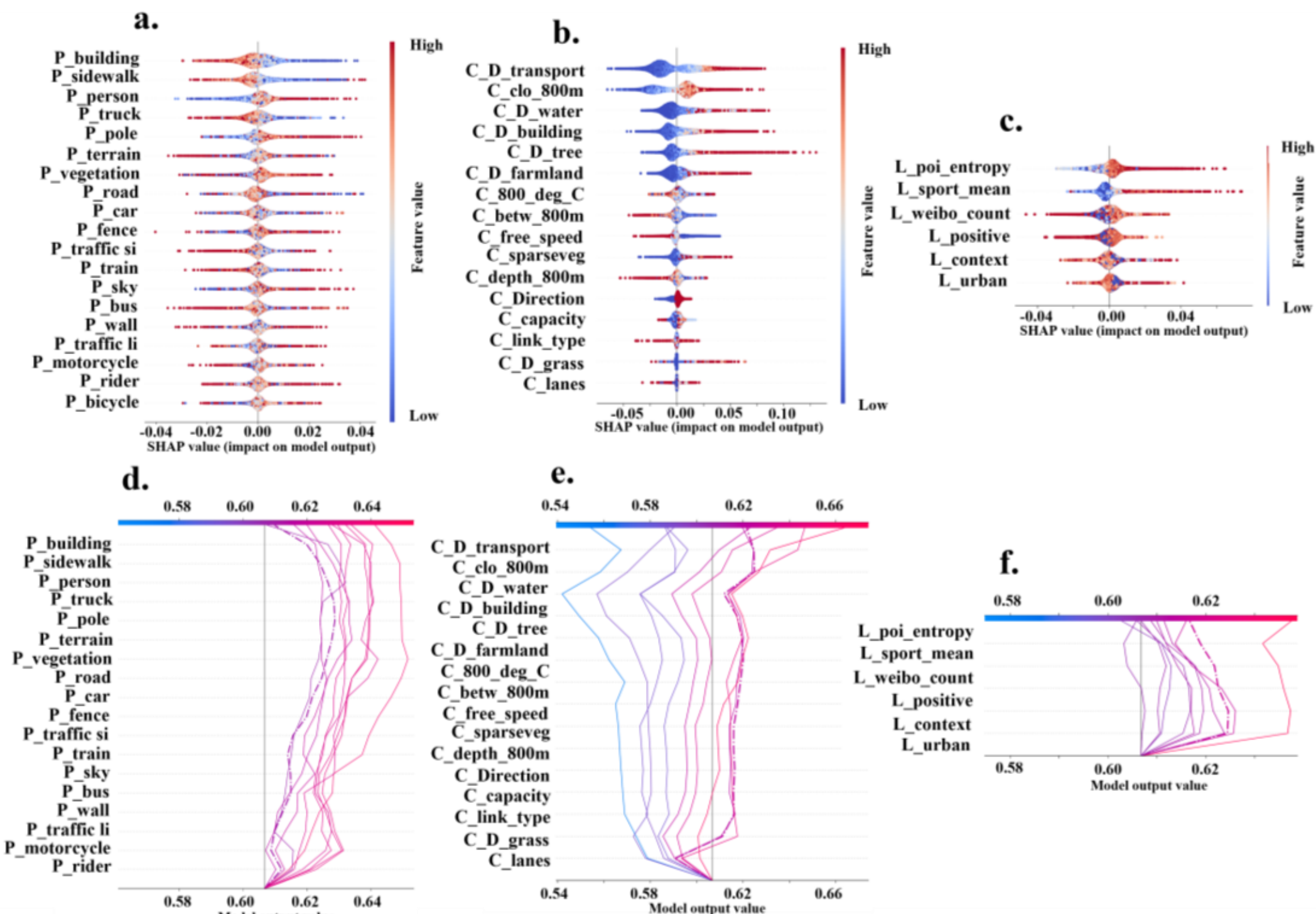

**Figure 5 SHAP summary for (a.) Perceived Space variables (P); (b.) Conceived Space variables (C); (c.) Lived Space variables (L), and decision path for (d.) Perceived Space variables (P); (e.) Conceived Space variables (C); (f.) Lived Space variables (L)**

Figure 5 further disaggregates these dimensions, revealing how specific features shape movement through distinct mechanisms. Within the dominant Conceived dimension, variables such as transport density ($C_D_{\text{transport}}$), closeness centrality ($C_{\text{clo_800m}}$), and proximity to water bodies ($C_D_{\text{water}}$) display consistent, positive decision-path trajectories. These linear associations suggest that higher functional connectivity and spatial accessibility robustly elevate predicted exercise intensity. Consequently, the urban network structure plays a foundational role, acting as the primary enabler that determines the baseline capacity for street-level movement

In contrast, the Perceived space features derived from street-level imagery introduce more complex, non-linear effects. Variables such as building enclosure ($P_{\text{building}}$) and sidewalk proportion ($P_{\text{sidewalk}}$) exhibit both positive and negative contributions depending on their intensity. The decision paths reveal specific thresholds beyond which additional built form or infrastructure density begins to yield diminishing returns, reducing exercise potential rather than enhancing it. Similar saturation effects appear for terrain features ($P_{\text{terrain}}$). These patterns suggest that sensory environments support movement only up to optimal levels of enclosure and visual complexity, a finding consistent with established research on legibility and human-scale comfort where moderation often outperforms excess

Finally, features of Lived space contribute more modestly to the overall variance yet exert strong influence at high quantiles. Indicators such as total Weibo activity ($L_{\text{weibo_count}}$), positive sentiment ($L_{\text{positive_mean}}$), and sport-related discourse ($L_{\text{sport_mean}}$) show sharp increases in their effect sizes once social activity or emotional positivity reaches a critical mass. This indicates a non-linear amplification mechanism where existing social vibrancy catalyzes further movement. By contrast, the diversity of amenities ($L_{\text{poi_entropy300m}}$) remains a relatively stable predictor. Taken together, these results demonstrate a complementary triadic logic: conceived structure provides the necessary access, perceived quality determines the sensory comfort threshold, and lived experience acts as a dynamic amplifier of activity once the physical conditions are met.

### 4.2.2 District-Level Heterogeneity and Spatial Signatures

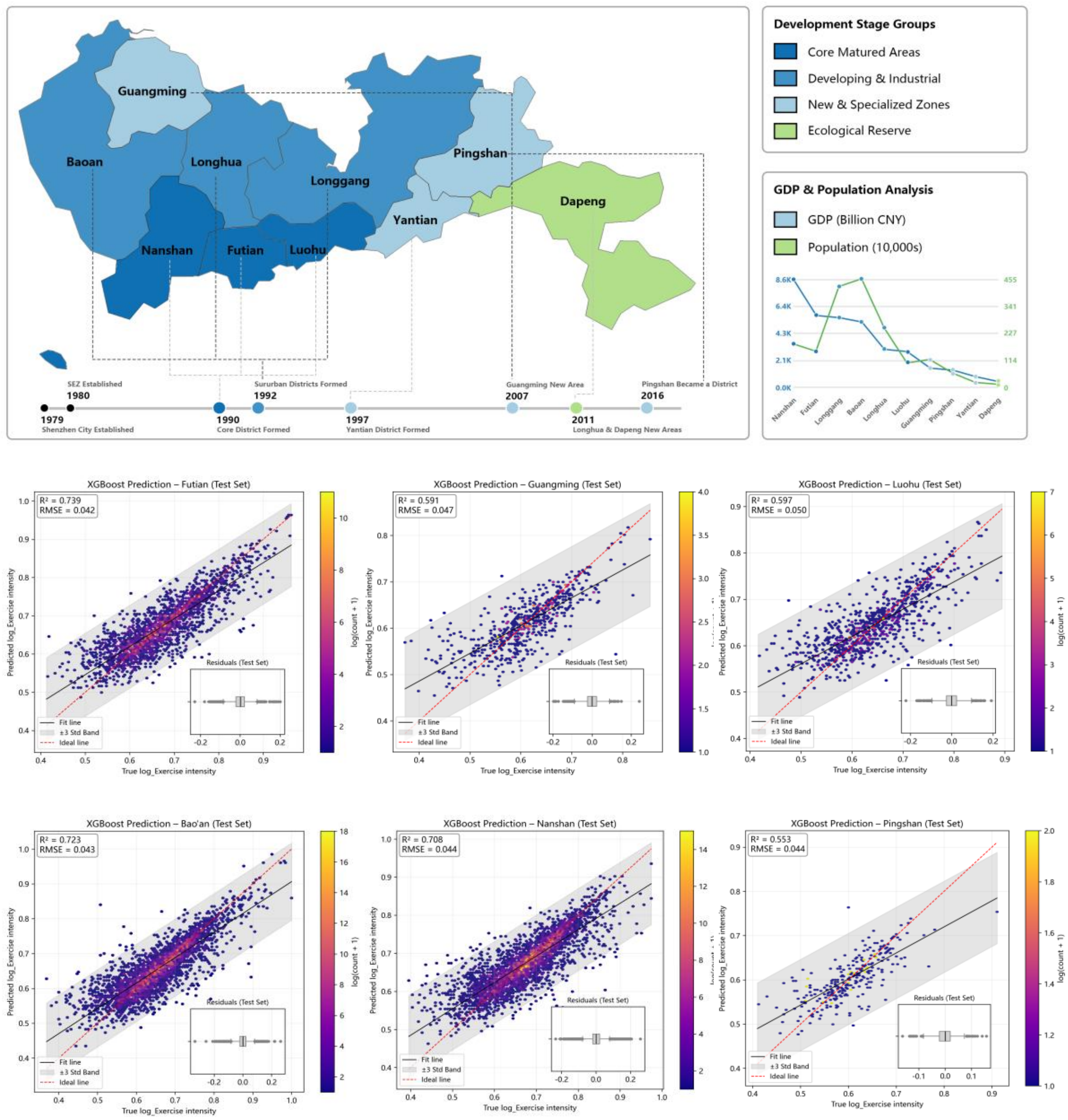

**Figure 6 District-specific XGBoost model performance, development stage context, and demographic–economic profiles across six Shenzhen districts. The upper panel shows Shenzhen's administrative districts grouped by development stage (core matured, developing/industrial, new and specialized zones, ecological reserve) with a timeline of key district formations (1979–2016). The line graph depicts 2022–2023 GDP (billion CNY) and population (10,000s) for each district, illustrating marked differences in economic scale and demographic density. The lower scatter plots display predicted versus observed street-level exercise intensity from district-specific XGBoost models, with 95 % confidence bands and residual histograms.**

While citywide analysis reveals general laws, it risks masking local variations rooted in urban development history. To disentangle this spatial heterogeneity, we conducted a subregional decomposition across six representative districts spanning a developmental gradient: Futian and Luohu (mature institutional cores), Bao'an and Nanshan (evolving innovation-industrial zones), and Guangming and Pingshan (emerging peripheral new towns). By synthesizing district-specific XGBoost scatterplots with development timelines and socioeconomic profiles (Figure 6), we contextualized how varying governance regimes and urban form maturity shape localized deprivation logics.

The modeling results reveal a striking correlation between urban maturity and predictive capacity. As shown in Figure 6, models fitted to mature cores exhibit strong predictive power, with $R^2$ scores reaching 0.739 in Futian. Conversely, the newly established districts of Guangming and Pingshan yield notably weaker fits ($R^2$ approx. 0.55). This divergence suggests that in established cores, the high density of amenities and legible street networks creates a deterministic environment where spatial features reliably explain behavior. In contrast, the lower predictability in peripheral new towns likely reflects a condition of "spatial immaturity," where stochastic user behavior prevails due to weak spatial legibility and a fragmented public realm (Lynch, 1960; Batty, 2013).

To decode these local mechanisms, we refined our analytical resolution by disaggregating the Conceived dimension into $C_{structure}$ (road-intrinsic connectivity) and $C_{landuse}$ (surrounding functional context). The resulting SHAP composition rose diagrams (Figure 7) visualize distinct spatial signatures that evolve with the district's development stage. In the mature cores of Futian and Luohu, the profiles exhibit a balanced triadic synergy, with a mix of structure, land use, and perceived quality complemented by a visible contribution from Lived space. In Luohu, lived space reaches nearly 15% of the predictive power, aligning with its long-established urban fabric where sustained economic density has produced a vibrant social presence.

The innovation-industrial zones display a diverging trajectory. Nanshan shows high Perceived-space contributions alongside a robust Structural base, reflecting a "co-evolution" where infrastructure planning and streetscape beautification have advanced in tandem. In contrast, Bao'an displays a mismatch where Land-use and Perceived factors dominate but lived contribution is weaker, reflecting its status as a maturing mix where large-scale planning has established the skeleton but human-scale social activation lags behind16. Finally, in the peripheral new towns of Guangming and Pingshan, the spatial logic is overwhelmingly dominated by macro-planning ($C_{landuse}$), suggesting that activity patterns remain heavily structured by coarse zoning rather than fine-grained human-scale affordances. These observed asymmetries confirm that deprivation is not merely spatial but chronological, motivating the diagnostic typology that follows.

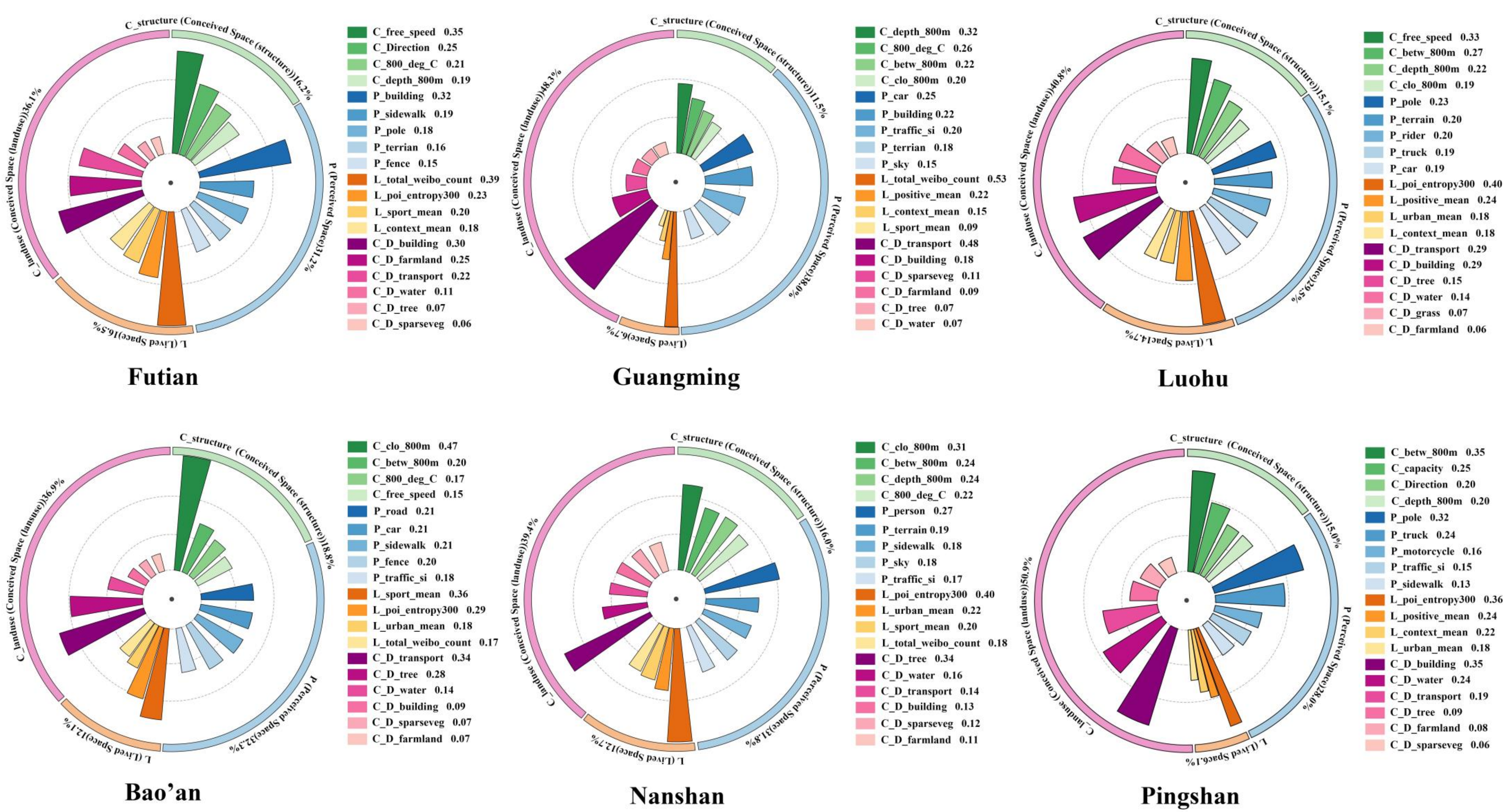

**Figure 7 SHAP-based triadic contribution rose diagrams for six representative districts.**

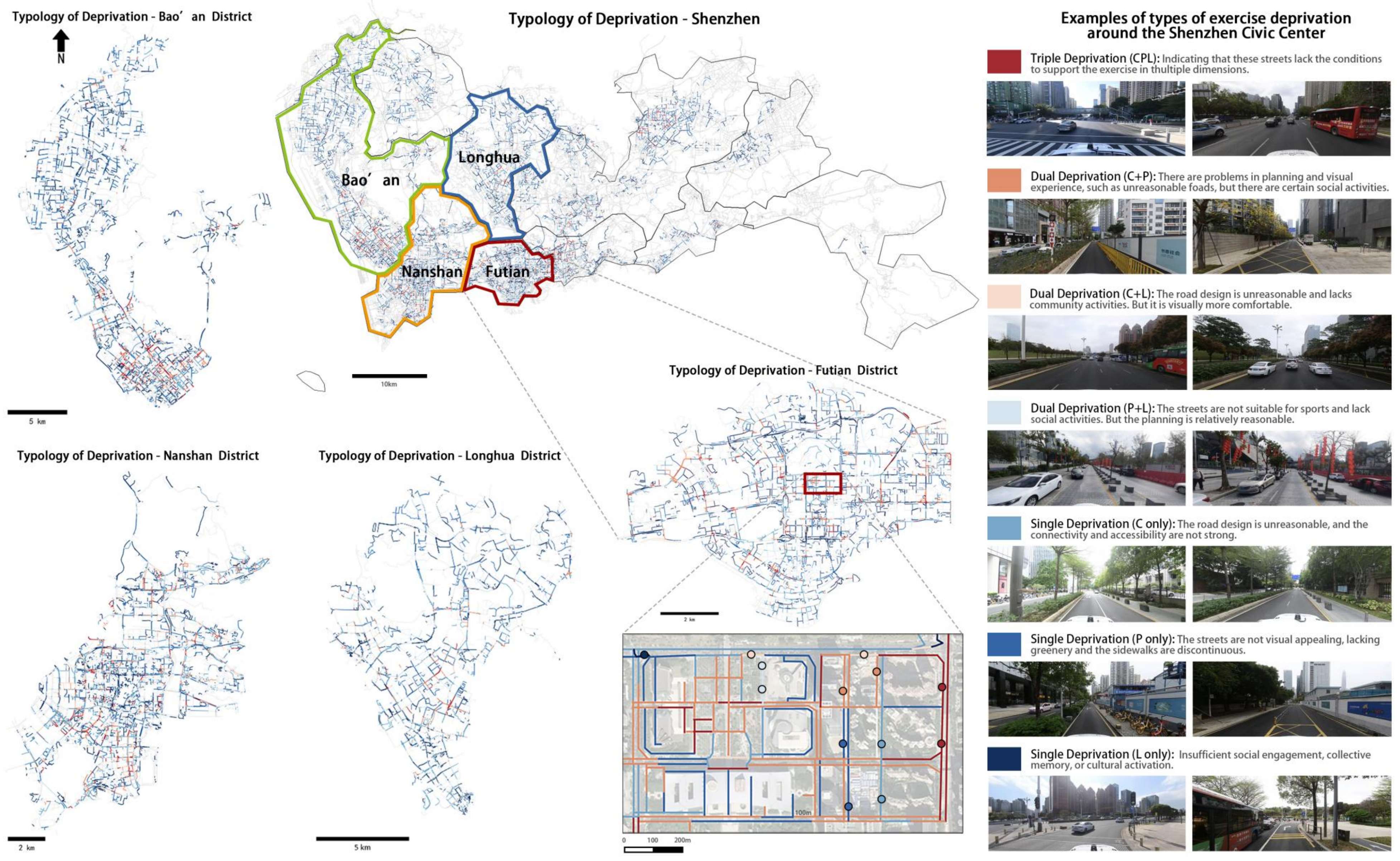

**Figure 8 Typology of street-level exercise deprivation across Shenzhen and representative districts. The left and center panels map the seven deprivation modes (C-only, P-only, L-only, CP, CL, PL, CPL) at the street-segment level for Shenzhen as a whole and for four representative districts (Futian, Nanshan, Bao'an, and Longhua). The inset zooms into a cluster near the Shenzhen Civic Center to illustrate fine-grained patterns. The right-hand column shows ground photographs of typical streets corresponding to each deprivation type, from triple deprivation (CPL) to single-domain deprivation, clarifying how conceived, perceived, and lived deficits manifest visually and spatially.**

### 4.3 A Diagnostic Typology of Exercise Deprivation

To translate these triadic insights into a street-level diagnosis, we synthesized the SHAP-derived contributions into a spatially explicit typology. By applying a relative threshold to the Conceived, Perceived, and Lived dimensions, each road segment was classified into one of seven mutually exclusive types: triple deprivation (CPL), dual deprivation (CP, CL, PL), or single-domain deprivation (C, P, or L only). Figure 8 maps these resulting typologies across the city and within four representative districts—Futian, Nanshan, Bao'an, and Longhua—alongside ground-level photographs illustrating each mode. When interpreted in conjunction with the development timelines and socioeconomic profiles presented in Figure 6, this typology provides a multidimensional context for understanding how specific deprivation patterns emerge under distinct urban paradigms.

In the mature administrative core of Futian, which is characterized by high GDP and population density, the analysis reveals a prevalence of L-only and PL deprivation concentrated around civic facilities and arterial roads. This indicates that while the physical planning infrastructure and basic visual qualities are adequate, social activation and everyday use often lag behind. This phenomenon echoes the "saturation effect" noted in Section 4.2, suggesting that in highly developed cores, the primary barrier is no longer structural access but the fostering of vibrant social practice.

Conversely, Nanshan, an innovation district developed earlier in Shenzhen's history, displays a different spatial logic. Here, P-only and CP patterns predominate, reflecting a condition where sound infrastructure exists but is compromised by perceptual barriers such as visual clutter, weak shading, or pedestrian discontinuities. This finding confirms that even in high-growth, mixed-use zones, the lack of sensory refinement remains a critical bottleneck that limits the comfort and sustainability of active use.

Moving to the transitional districts, Bao'an exhibits a complex spatial mosaic. Its older industrial corridors are characterized by C-only and CP deprivation, pointing to lingering deficiencies in both network structure and visual quality. In contrast, its newer housing clusters display sporadic L-only deprivation. This latter pattern mirrors the situation in Longhua, a rapidly developing "new town," where physical connectivity has outpaced social and experiential activation. Longhua's strong prevalence of C-only and L-only modes underlines a significant temporal lag: the physical city has been built, but the "lived city" of social memory and cultural activation has yet to fully emerge.

These subregional variations affirm a central theoretical argument of this study: deprivation is not merely a function of absence, but a condition of imbalance. Streets are deprived not simply because they lack design or activity, but because what is planned, what is seen, and what is lived fail to align—a core insight derived from Lefebvre's triadic theory (Lefebvre, 2012; Simonsen, 2007). This analysis extends the reading of theory into practice. By converting abstract SHAP values into structured deprivation types, we create a diagnostic language for interpreting exactly how and why street environments fail to support healthful exercise.

Crucially, this typology serves not only an explanatory function but also a prescriptive one. In this respect, the framework closely aligns with the central argument advanced by Mei-Po Kwan and Tim Schwanen (2009a), who called for revitalizing the critical consciousness of quantitative geography. Their proposal advocates moving beyond the binary opposition between critique and quantification and positions quantitative geographical analysis as a powerful instrument for challenging social and global injustices while contributing to progressive social and political transformation. Each deprivation type implies a distinct intervention pathway: structural redesign targeting conceived space (C), environmental articulation to enhance perceived space (P), or social animation focusing on lived space (L). This approach enables spatial justice strategies that are both place-sensitive and behaviorally grounded (Soja, 2010). Taken together, the analysis demonstrates how a theory-guided machine learning framework can uncover latent inequalities embedded in everyday street life, illustrating that the spatial determinants of urban exercise vary not only in strength but in kind across development regimes. In mature districts, behavioral saturation demands perceptual and social refinement; in innovation corridors, visibility and imageability shape engagement; and in new towns, spatial justice begins not with more infrastructure, but with more meaning.

## 4.4 From Diagnosis to Intervention: Spatial Mismatch Analysis

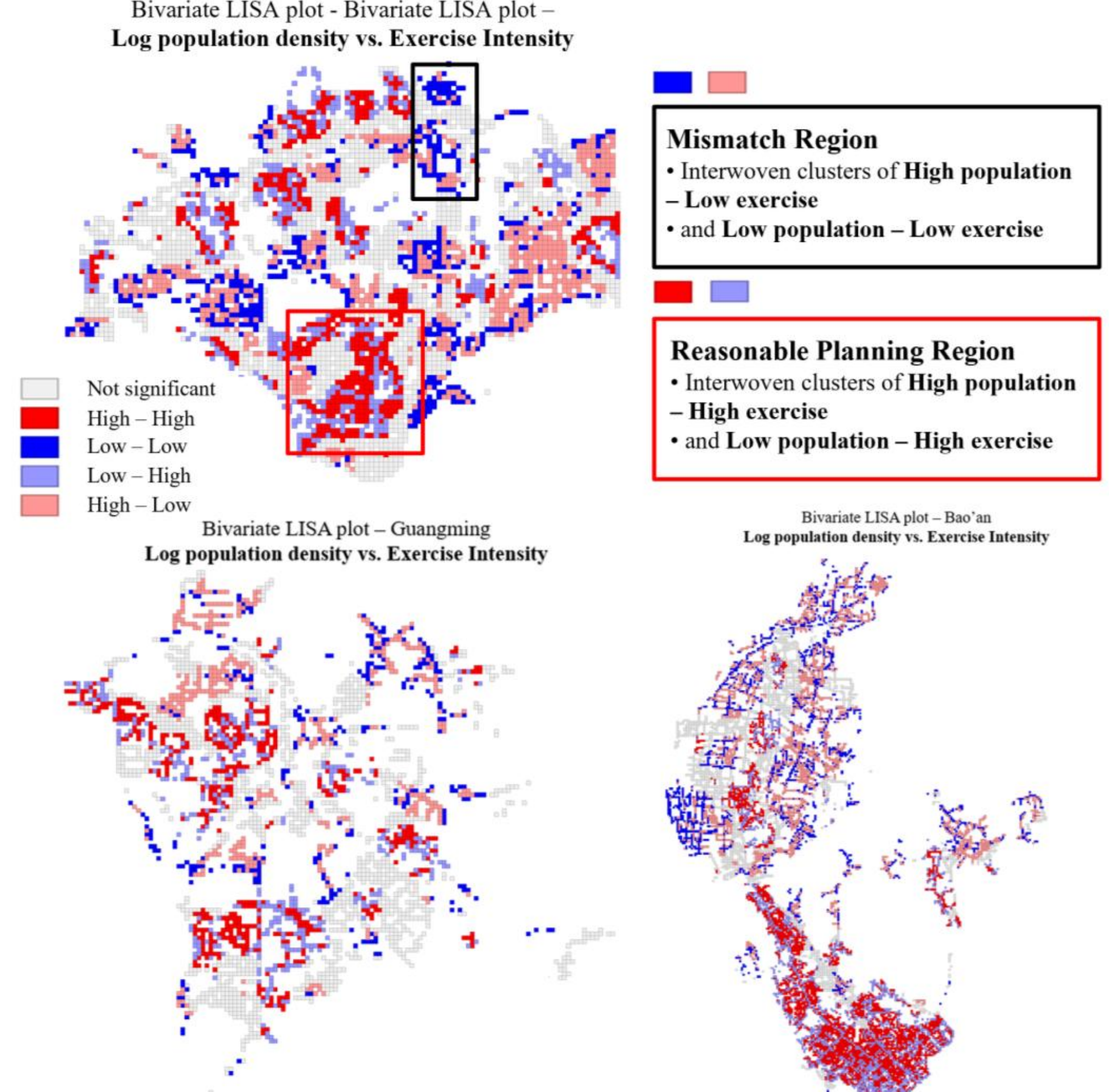

**Figure 9 Identifying Spatial Mismatches: Bivariate LISA Analysis of Population Demand vs. Exercise Support. Maps depict local spatial associations (LISA) between resident population density (demand) and predicted street-level exercise intensity (supply) across Shenzhen and representative districts. (Red & Dark Blue) Zones of Congruence: Areas where supply matches demand (High-High or Low-Low), indicating balanced planning. (Pink & Light Blue) Zones of Mismatch: Critical areas of misalignment, particularly the High-Demand/Low-Support clusters (Pink/Light Red), which pinpoint priority locations where demographic pressure exceeds current spatial capacity, necessitating targeted intervention.**

In complex urban systems, not all spatial deprivation constitutes an equity failure. Low-support segments located in low-demand zones may represent appropriate resource allocation, whereas similar deficits in high-demand settings signal a critical gap in service provision (Xie et al., 2023). To systematically identify these critical gaps, we transitioned from segment-based diagnostics to a grid-level bivariate LISA analysis, juxtaposing log population density (demand) against predicted exercise supportiveness (supply). This scalar shift allows us to capture the broader spatial relationship between demographic pressure and environmental capacity.

As visualized in Figure 9, three characteristic spatial configurations emerge from this bivariate reading. At one end lie zones of "spatial congruence", where high population density coincides with high support (High-High) or low population aligns with low support (Low-Low); in these areas, supply and demand are broadly balanced. Conversely, clusters of "spatial mismatch" emerge where high demand coincides with low exercise support (High-Low) or vice versa, revealing latent inequities in infrastructure provision. Between these poles lie transitional areas without clear clustering. This continuum underscores that deprivation is not a binary absence but a relational misalignment between where people live and where the city supports their movement.

This spatial reading reveals distinct typologies of injustice. By overlaying population demand onto exercise supportiveness, we highlight how certain districts, particularly in rapidly transforming areas, contain extensive High-Demand/Low-Support (HL) clusters. These areas represent "latent spatial injustices" where physical conditions underdeliver relative to human need. In Futian, for example, mismatch zones cluster around older arterial corridors where density outstrips aging infrastructure; in Bao'an and Guangming, they appear along industrial–residential edges, reflecting a lag in public-realm investment relative to population influx. These findings confirm that the planning challenge is not merely providing infrastructure but ensuring its distributional alignment with everyday life.

Targeting these HL mismatch clusters transforms the analysis from diagnosis to remediation. While the LISA analysis identifies the location of the mismatch, the SHAP-derived typology identifies the nature of the remedy. By cross-referencing HL grids with the deprivation modes identified in Section 4.3, we can define precise intervention pathways. For instance, C-heavy HL grids necessitate structural redesign to improve connectivity; P-heavy HL grids require environmental articulation such as greening or facade improvements; and L-heavy HL grids demand community activation strategies. This integration of grid-level targeting and segment-level diagnosis transforms abstract model outputs into actionable planning priorities. Detailed intervention simulations and quantitative assessments of these scenarios are provided in the Supplementary Materials.

### 4.5 Limitations and Scope of Inference

This study provides a diagnostic reading of street-based exercise, yet its inferences must be contextualized within specific interpretative boundaries. A primary analytical challenge lies in the translation of Lefebvre's spatial triad into a quantitative model. Converting the fluid, recursive complexity of "conceived," "perceived," and "lived" spaces into discrete variable sets entails a necessary trade-off that prioritizes planning operability over philosophical exhaustiveness. This reductionism extends to the underlying datasets, which act as high-resolution proxies rather than comprehensive measures. For instance, app-based trajectories reflect the preferences of digitally connected, fitness-oriented cohorts rather than the full demographic spectrum, while street-view imagery captures visual composition but may miss the embodied sensory nuances of the urban experience.

Methodologically, it is important to note that SHAP values describe statistical associations rather than definitive causal mechanisms. Given the inherent correlations between urban features, the resulting typology functions best as a structured heuristic, a diagnostic guide for identifying likely deficits, rather than rigid causal accounting. Finally, as these findings are shaped by Shenzhen's specific morphological and institutional context, applying this workflow to other cities would require calibrating feature sets to local norms. Future research could further strengthen this approach by integrating longitudinal data or participatory validation, thereby bridging the gap between algorithmic diagnosis and on-the-ground behavioral change.

# 5 Conclusion

This study reconciles the predictive capabilities of urban analytics with the interpretive depth of critical-geographical theory to diagnose street-level exercise deprivation. By employing Henri Lefebvre's spatial triad as a lens rather than a rigid taxonomy, we move beyond "black-box" indices to reveal how planning form, sensory quality, and social practice distinctively constrain active living. Empirical analysis in Shenzhen yields a fundamental insight: deprivation arises less from the absolute absence of resources than from a relational misalignment among these dimensions. While conceived attributes provide the structural baseline, local mechanisms vary significantly, ranging from behavioral saturation in mature cores to structural coarseness in new towns. This demonstrates that streets often fail to support activity not because they lack infrastructure, but because the planned, perceived, and lived environments are out of sync.

Practically, the resulting diagnostic typology supports a paradigm shift from blanket supply to precision urban repair. By tailoring interventions to specific deprivation profiles, which may require structural redesign, environmental articulation, or community activation, planners can direct resources where they are most effective. As cities worldwide grapple with densification and health equity, this triad-informed approach offers a robust pathway for reclaiming street space as truly inclusive health infrastructure.

## Appendix A: Prompt Design for LLM-assisted Urban Perception Scoring

To analyze street-scale social media texts in terms of urban perception and experiential context, we developed an LLM-assisted semantic scoring approach using a structured prompt. The core prompt used in the analysis is presented below.

**System Prompt：**

You are an expert in urban space perception analysis. Please evaluate each text along the following nine dimensions, using a score from 0 to 1 with two decimal places, and return the results in standard JSON format.

**User Prompt Template：**

The text is as follows:
{text}
Please score the following microblog texts across multiple affective and contextual dimensions, using a scale from 0 to 1 with two decimal places. The dimensions are defined as follows:
1. Urban: the extent to which the text involves evaluation or description of urban space, streets, landscape, environment, urban renewal, or related urban elements.
2. positive: the extent to which the text expresses positive or energetic emotions (e.g., happiness, satisfaction, vitality).
3. relax: the extent to which the text conveys relaxation, calmness, or comfort (e.g., slow-paced life, quietness, ease).4. explore: yearning/exploring (want to go, curious, etc.)
4. explore: the extent to which the text expresses curiosity, interest, or a desire to explore or visit.
5. social: the extent to which the text reflects social interaction, gathering, companionship, or collective activity.
6. sport: the extent to which the text is related to sport or physical activity.
7. safe_walk: the extent to which the text refers to safety or walkability, such as road conditions, lighting, accessibility, or walking experience. Use 0.50 if this aspect is not mentioned.
8. context: the extent to which the text reflects the immediate geographical context or present-moment experience, that is, the consistency between the description and the real location.
The microblog texts are:
{text_block}

Please return the results in the following format only, without any explanation. Each text should correspond to one JSON object:

{
 "1": {"urban": 0.85, "positive": 0.76, "relax": 0.42, "explore": 0.67, "social": 0.31, "sport": 0.05, "safe_walk": 0.50, "context": 0.92},

```
  "2": {...},
  ...
  "10": {...}
}"""
```

### Validation of the LLM-assisted semantic assessment

To ensure the reliability of the LLM-assisted semantic evaluation introduced in Section 2.3, a human–machine consistency test was conducted using a sample of 50 social-media posts representing diverse spatial and contextual situations. Two trained researchers with urban-planning backgrounds (Reviewer A and Reviewer B) independently rated each post on a five-level exercise-supportiveness scale consistent with the automated model output.

The inter-rater and human–machine agreements were assessed using both the overall agreement rate (Acc) and Cohen's κ, defined as

$$Acc = \frac{1}{n}\sum_{i=1}^{n} I\,(r_{1i} = r_{2i}), \quad \kappa = \frac{p_o - p_e}{1 - p_e}$$

where $r_{1i}$ and $r_{2i}$ are the paired ratings for the $i^{th}$ item, $p_o$ is the observed proportion of agreement, and $p_e$ is the expected agreement by chance.

The analysis showed that the agreement between the two human coders (A–B) reached an Acc of 0.88 and a κ of 0.81. The agreement between human and model ratings was similarly high, with A–LLM = 0.85 (κ = 0.77) and B–LLM = 0.86 (κ = 0.78). All κ values fall within the *substantial agreement* range (0.61–0.80) according to Landis and Koch (1977). Minor inconsistencies mainly occurred for neutral-level posts (score 3), typically reflecting ambiguous tone or mixed contextual signals.

These results confirm that the LLM-assisted interpretation aligns closely with expert human judgment, supporting its use as a scalable and valid tool for large-scale urban perception analysis.

# Appendix B: Citywide SHAP Analysis - Variable Dependency

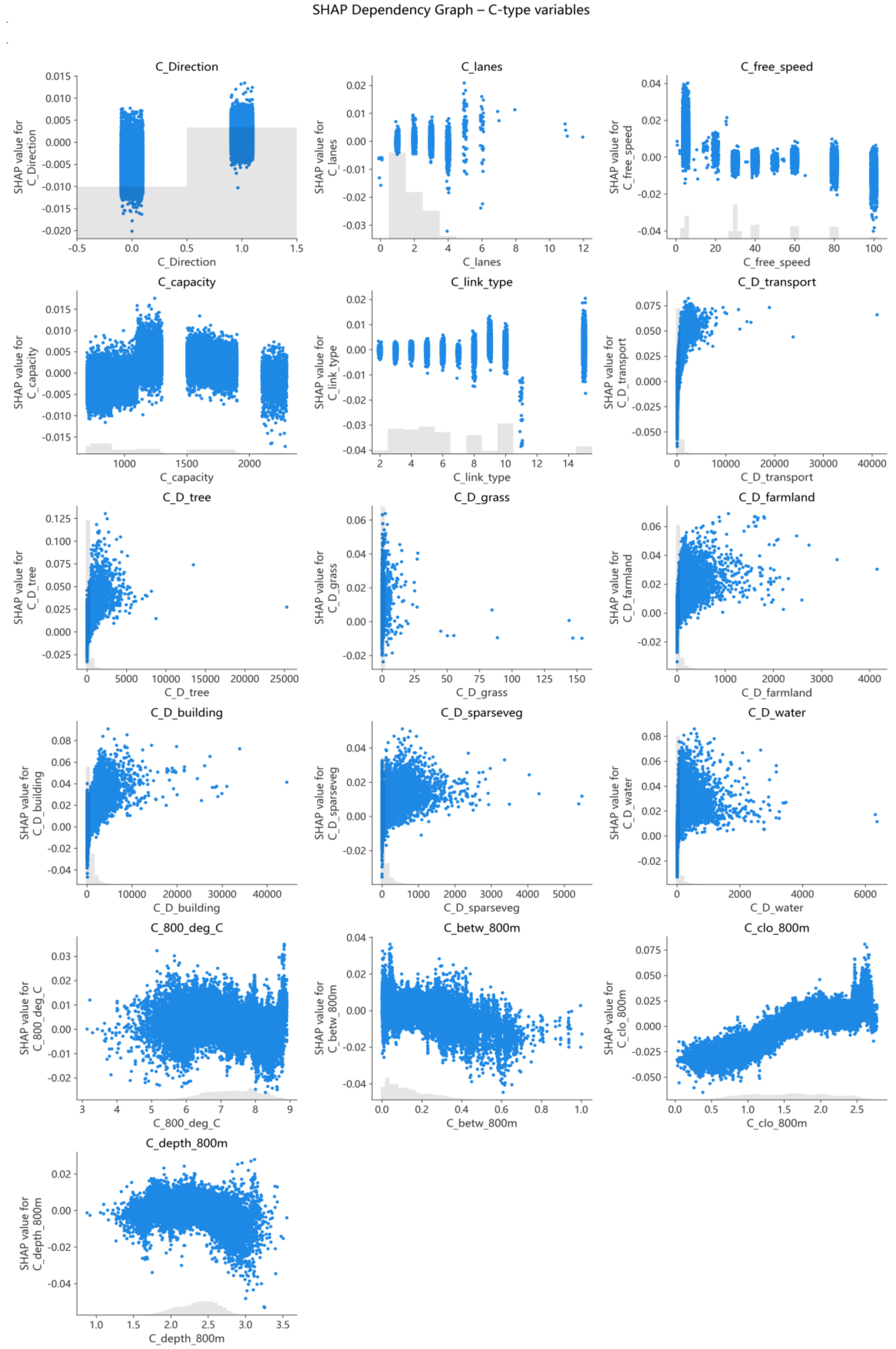

**Figure B.1 SHAP dependency graph of C-variables**

SHAP Dependency Graph – L-type variables

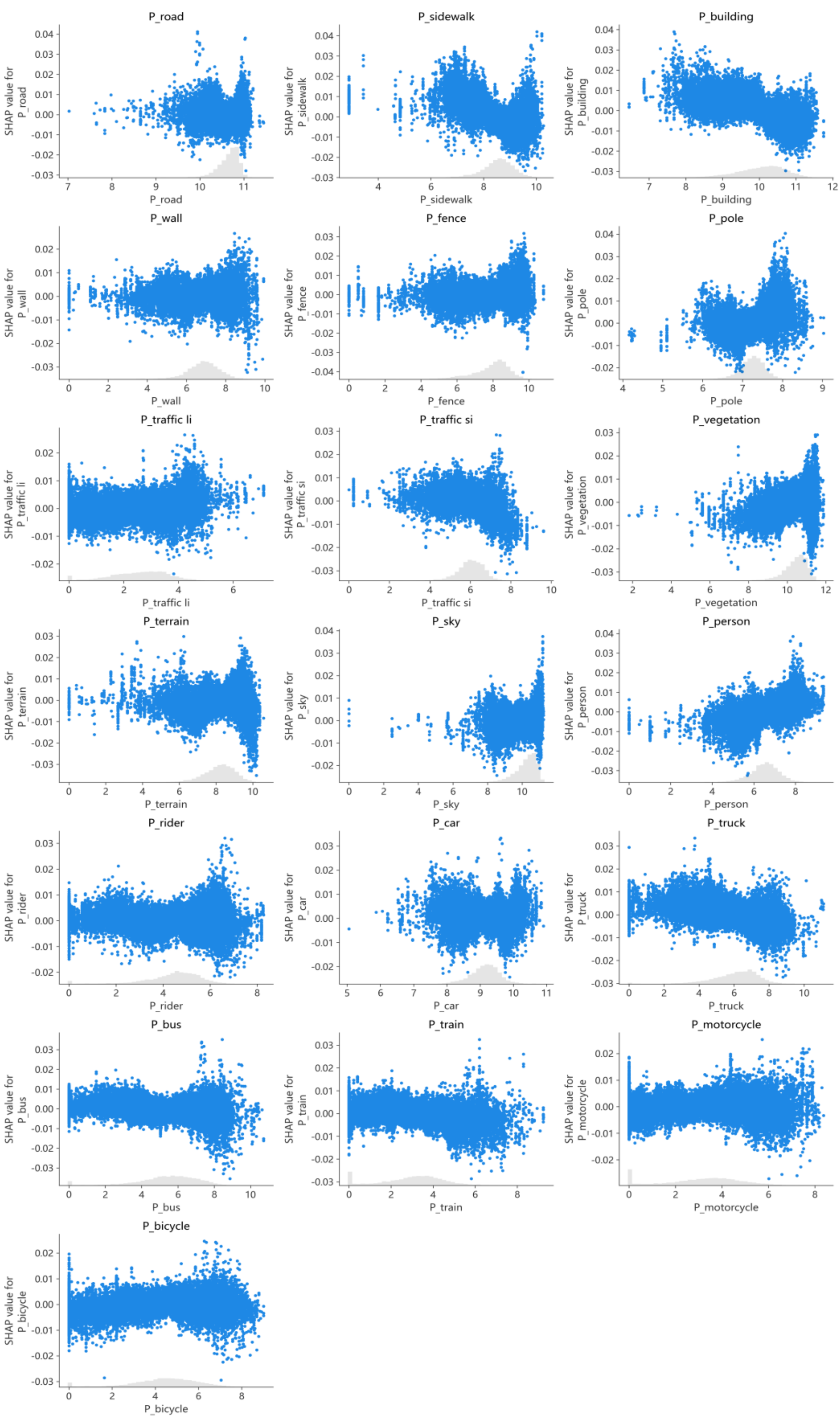

**Figure B.2 SHAP dependency graph of P-variables**

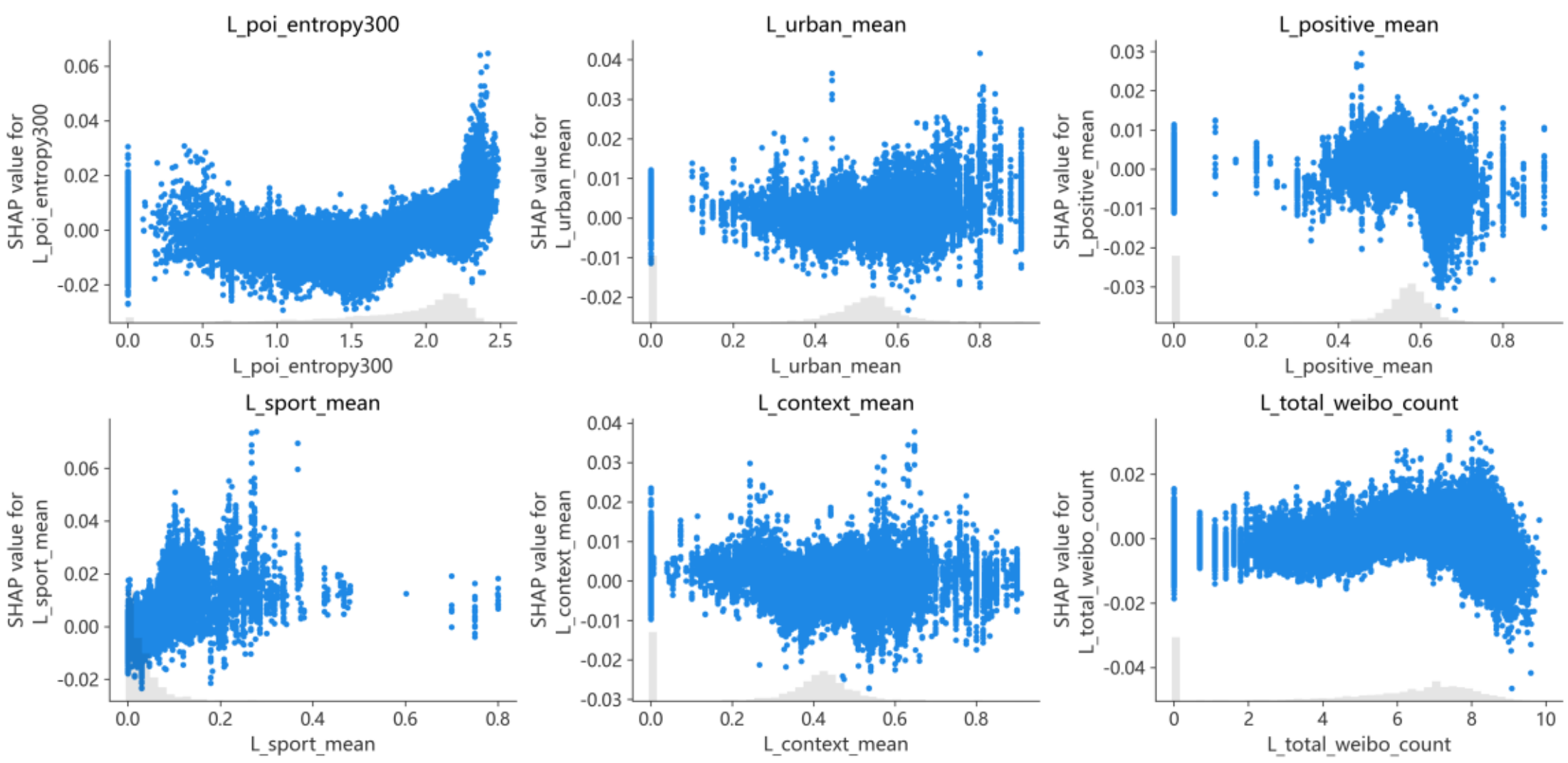

**Figure B.3 SHAP dependency graph of L-variables**

## Appendix C: Hyperparameter Search Space and Final Configuration

To ensure robust model performance and avoid overfitting, we conducted a randomized grid search across a carefully selected parameter space for each regression algorithm. The search employed 10-fold cross-validation and was performed separately for each model using the training set.

**Table C.1 Random Search Parameter Space**

| Model | Parameter | Search Range |
|---|---|---|
| XGBoost | max_depth | [3, 4, 5, 6, 8] |
| | learning_rate | [0.01, 0.05, 0.1, 0.2] |
| | subsample | [0.6, 0.8, 1.0] |
| | colsample_bytree | [0.6, 0.8, 1.0] |
| | gamma | [0, 0.1, 0.5, 1] |
| RandomForest | max_depth | [4, 6, 8,] |
| | n_estimators | [100, 150, 200, 300] |
| | max_features | ['auto', 'sqrt', 'log2'] |
| LightGBM | max_depth | [4, 6, 8] |
| | learning_rate | [0.01, 0.05, 0.1] |
| | num_leaves | [31, 63, 127] |
| | feature_fraction | [0.6, 0.8, 1.0] |

**Table C.2 Optimal Hyperparameters for XGBoost Model**

| Parameter | Value | Description |
|---|---|---|
| objective | 'reg:squarederror' | Loss function for regression tasks |
| n_estimators | 200 | Number of boosting rounds |
| max_depth | 8 | Maximum depth of each decision tree |
| learning_rate | 0.05 | Step size shrinkage |
| subsample | 0.8 | Fraction of samples used per tree |
| colsample_bytree | 0.8 | Fraction of features used per tree |
| gamma | 0.1 | Minimum loss reduction to make a split |
| random_state | 42 | Seed for reproducibility |

# Appendix D: SHAP-Informed Intervention Simulation

**Table 3 Summary of intervention simulations across Futian, Bao’an, and Guangming, including intervention type, variable count, and performance gains.**

<table>
<tr><th></th><th>Type of intervention</th><th>Intensity of intervention</th><th>Number of variables</th><th>Exercise Improves</th></tr>
<tr><td rowspan="10"><b>Futian</b></td><td><i>C</i></td><td>20%</td><td>5</td><td>3.71%</td></tr>
<tr><td><i>P</i></td><td>20%</td><td>5</td><td>2.54%</td></tr>
<tr><td><i>L</i></td><td>20%</td><td>5</td><td>3.21%</td></tr>
<tr><td><i>C+P</i></td><td>20%</td><td>5</td><td>6.32%</td></tr>
<tr><td><i>C+L</i></td><td>20%</td><td>5</td><td>6.80%</td></tr>
<tr><td><i>L+P</i></td><td>20%</td><td>5</td><td>3.75%</td></tr>
<tr><td><i>C+P+L</i></td><td>20%</td><td>5</td><td>7.00%</td></tr>
<tr><td><i>C+P+L</i></td><td>20%</td><td>10</td><td>8.05%</td></tr>
<tr><td><i>C+P+L</i></td><td>30%</td><td>10</td><td>8.17%</td></tr>
<tr><td><i>C+P+L</i></td><td>30%</td><td>15</td><td>10.02%</td></tr>
<tr><td colspan="5"><b>Top 10 variables:</b> 'L_total_weibo_count', 'P_sky', 'C_free_speed', 'L_positive_mean', 'P_building', 'C_depth_800m', 'P_fence', 'C_D_tree', 'P_car', 'L_urban_mean'</td></tr>
<tr><td rowspan="10"><b>Bao’an</b></td><td><i>C</i></td><td>20%</td><td>5</td><td>1.85%</td></tr>
<tr><td><i>P</i></td><td>20%</td><td>5</td><td>5.67%</td></tr>
<tr><td><i>L</i></td><td>20%</td><td>5</td><td>0.77%</td></tr>
<tr><td><i>C+P</i></td><td>20%</td><td>5</td><td>7.25%</td></tr>
<tr><td><i>C+L</i></td><td>20%</td><td>5</td><td>2.65%</td></tr>
<tr><td><i>L+P</i></td><td>20%</td><td>5</td><td>5.67%</td></tr>
<tr><td><i>C+P+L</i></td><td>20%</td><td>5</td><td>7.25%</td></tr>
<tr><td><i>C+P+L</i></td><td>20%</td><td>10</td><td>9.38%</td></tr>
<tr><td><i>C+P+L</i></td><td>30%</td><td>10</td><td>9.38%</td></tr>
<tr><td><i>C+P+L</i></td><td>30%</td><td>15</td><td>11.53%</td></tr>
<tr><td colspan="5"><b>Top 10 variables:</b> 'P_car', 'P_truck', 'C_free_speed', 'P_wall', 'P_road', 'P_sidewalk', 'P_building', 'C_D_sparseveg', 'L_poi_entropy300', 'P_bus'</td></tr>
<tr><td rowspan="10"><b>Guangming</b></td><td><i>C</i></td><td>20%</td><td>5</td><td>1.99%</td></tr>
<tr><td><i>P</i></td><td>20%</td><td>5</td><td>7.04%</td></tr>
<tr><td><i>L</i></td><td>20%</td><td>5</td><td>0.18%</td></tr>
<tr><td><i>C+P</i></td><td>20%</td><td>5</td><td>6.79%</td></tr>
<tr><td><i>C+L</i></td><td>20%</td><td>5</td><td>1.66%</td></tr>
<tr><td><i>L+P</i></td><td>20%</td><td>5</td><td>7.04%</td></tr>
<tr><td><i>C+P+L</i></td><td>20%</td><td>5</td><td>6.79%</td></tr>
<tr><td><i>C+P+L</i></td><td>20%</td><td>10</td><td>8.24%</td></tr>
<tr><td><i>C+P+L</i></td><td>30%</td><td>10</td><td>8.31%</td></tr>
<tr><td><i>C+P+L</i></td><td>30%</td><td>15</td><td>7.93%</td></tr>
<tr><td colspan="5"><b>Top 10 variables:</b> 'P_car', 'P_building', 'C_800_deg_C', 'P_traffic si', 'P_terrain', 'C_D_tree', 'P_fence', 'P_truck', 'P_road', 'C_clo_800m'</td></tr>
</table>

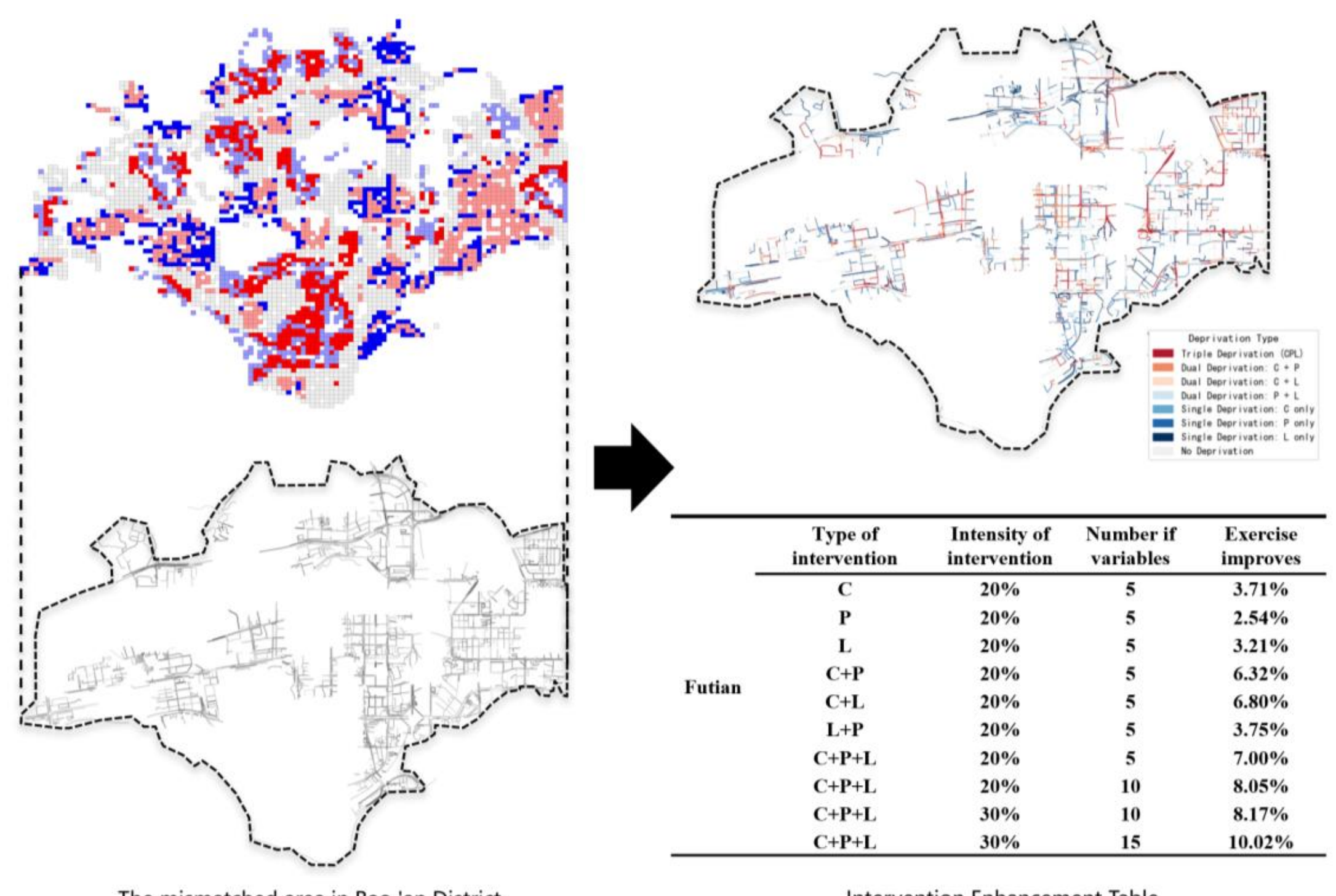

| | Type of intervention | Intensity of intervention | Number if variables | Exercise improves |
|---|---|---|---|---|
| Futian | C | 20% | 5 | 3.71% |
| | P | 20% | 5 | 2.54% |
| | L | 20% | 5 | 3.21% |
| | C+P | 20% | 5 | 6.32% |
| | C+L | 20% | 5 | 6.80% |
| | L+P | 20% | 5 | 3.75% |
| | C+P+L | 20% | 5 | 7.00% |
| | C+P+L | 20% | 10 | 8.05% |
| | C+P+L | 30% | 10 | 8.17% |
| | C+P+L | 30% | 15 | 10.02% |

**Figure D.1 SHAP-informed intervention simulation workflow for mismatched areas in Bao'an District. The left panel shows grid-level mismatch clusters where high population demand coincides with low exercise s $R^2$upport; the right panel maps the seven deprivation modes for the same area, overlaid with the simulated intervention enhancements. The accompanying table reports predicted exercise improvements under different intervention types (C, P, L and their combinations) and intensities.**

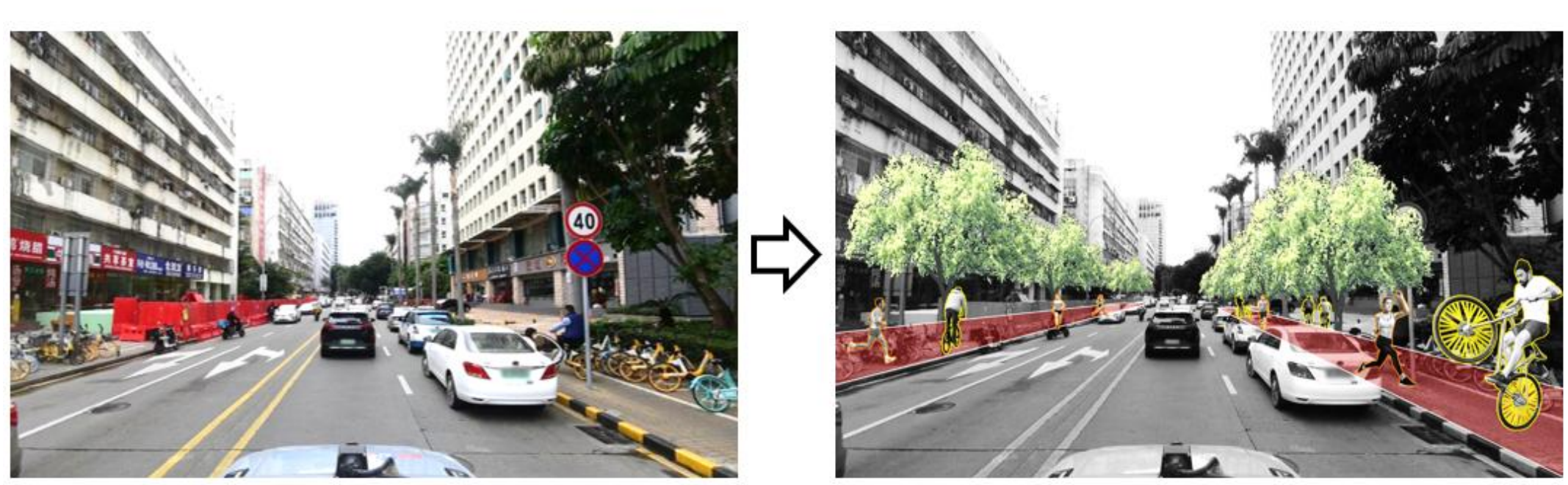

**Figure D.2 Street-level intervention scenarios illustrating how conceived (C), perceived (P), and lived (L) spaces can be improved. The left image shows the current condition, while the right visualizes hypothetical enhancements**

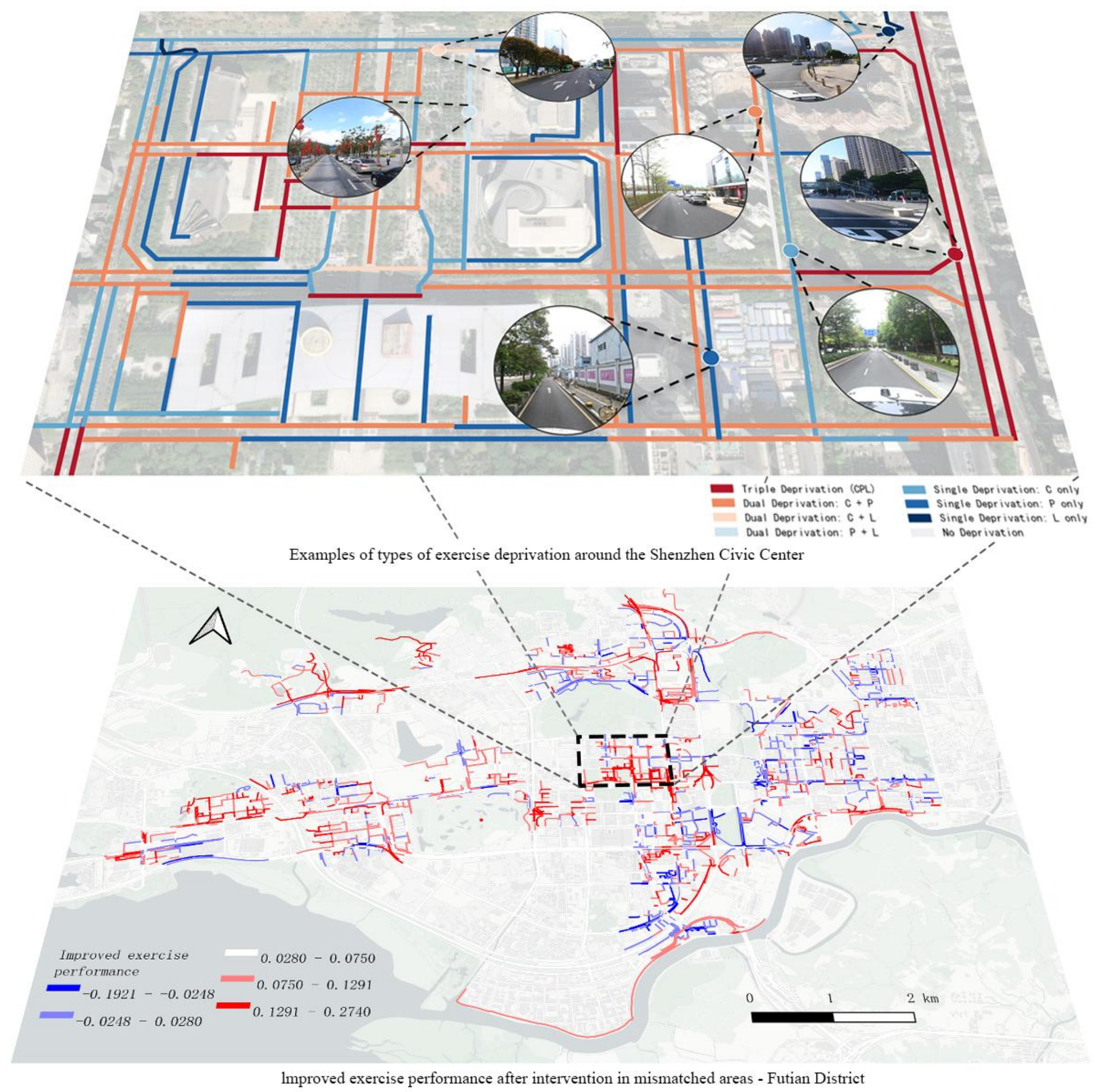

**Figure D.3 Examples of exercise deprivation types around the Shenzhen Civic Center (top) and the predicted improvements in exercise supportiveness after simulated interventions (bottom) in Futian District. Red and blue hues indicate the magnitude of predicted change**

We simulated SHAP-informed interventions by perturbing key explanatory variables in the top-ranked SHAP dimensions and recomputing predicted exercise scores within mismatch areas. Formally, for a given segment s, the adjusted prediction under a hypothetical intervention $\delta$ on feature $x_k$ is given by:

$$\hat{f}_s' = f(x_1, \ldots, x_k + \delta, \ldots, x_n)$$

Aggregated intervention effects were summarized by typology and region, resulting in a structured intervention enhancement table that specifies (1) which spatial domain (C/P/L) to target, (2) which specific features to improve (e.g., greening, sidewalk continuity, POI richness), and (3) the estimated benefit in predicted movement supportiveness.